# Controllable optical resonances and unidirectional scattering by core-shell nanoparticles

YiWei Dong[1]*, Yuanqing Yang[2]

**Abstract:** Nanoparticles supporting a distinct series of Mie resonances have enabled a new class of nanoantennas and provide efficient ways to manipulate light at the nanoscale. The ability to flexibly tune the optical resonances and scattering directionality are particularly essential for various applications ranging from biosensing to nanolasers.

In this paper, we investigate the core-shell nanoparticles that support both electric and magnetic Mie resonances and for the first time systematically reveal the mode evolution from a pure high-index dielectric nanosphere to a pure plasmonic one, where it has a sudden transition when core-shell ratio increases from 0.4 to 0.5. Furthermore, by engineering the electric and magnetic resonances, we demonstrate the unidirectional forward and backward scattering in such a system and reveal its tunability via geometric tuning.

Key words: Mie scattering, electrical and magnetic resonance, Kerker condition, unidirectional scattering

# Introduction

    Light-matter interaction is one of the most field in our scientific research and industry, because today we have relied on it heavily, from basic structure like our smartphone's screen[1], signal processing[2], aerial[3] to frontier field like solar panel[4] and nanolasers[5]. Starting from the seminal work of Mie[6], light scattering by small particles has attracted great amount of attention in many fields of

1. Department of Physics, Beijing Normal University, Beijing 100875, China
2. ASML Netherlands B.V. (Netherlands)

physics[7].At first, because Mie theory is complex and mostly suits for particles whose diameter is close to incident light's wavelength—in 1920s we couldn't make such devices to make a precise sphere's diameter can compare with visible light, nearly no one had interest about it. However, with the development of high precision machining, Mie theory came back to one of the central points for newborn semiconductor technology and discussing light-metal interaction. One important work was done by Kerker, who discovered unidirectional scattering for some kinds of particles in specific conditions using Mie theory[8-9]. After his work, the strong research value and commercial potential of unidirectional scattering has attracted many researches[10-11]. However, most researches are focused on electrical or magnetic resonance or Kerker conditions with a single, unchanged pure nanoparticle[12] or core-shell nanoparticle[13], not one we can control dynamically. But in the industry, with the development of nanofabrication technology, even core-shell particle, we can control its core-shell ratio to make a series of nanoparticles and choose one best suits for our purpose. There is a gap of systematically discovering the light-matter interaction properties for a series of nanoparticles with different core-shell ratio.

In this paper, we studied the relationship between ED (electric dipole), MD (magnetic dipole) resonance, Kerker conditions and the core-shell ratio of an Ag core-Si shell nanoparticle in a systematically and dynamically view, and found the evolution process of those properties from a pure semiconductor nanoparticle to a pure metal nanoparticle, both in analytical method. We found that the wavelength for ED and MD resonance change in different way and the wavelength for first and second Kerker conditions not only change, but also sometimes vanish.

# Results and discussion

## Isolated dielectric and plasmonic nanospheres

Using the basic Mie theory that suits pure nanoparticles, we start our analysis by investigating the optical resonances within homogeneous dielectric and plasmonic nanoparticles positioned in the vacuum. We set the radius of the sphere to 80nm and calculate their scattering response via analytical Mie theory [7] (See methods)

Here we set the refractive index of a dielectric nanoparticle to 3.5 (refer to Si)[14-15], with no imaginary parts because it's negligible.

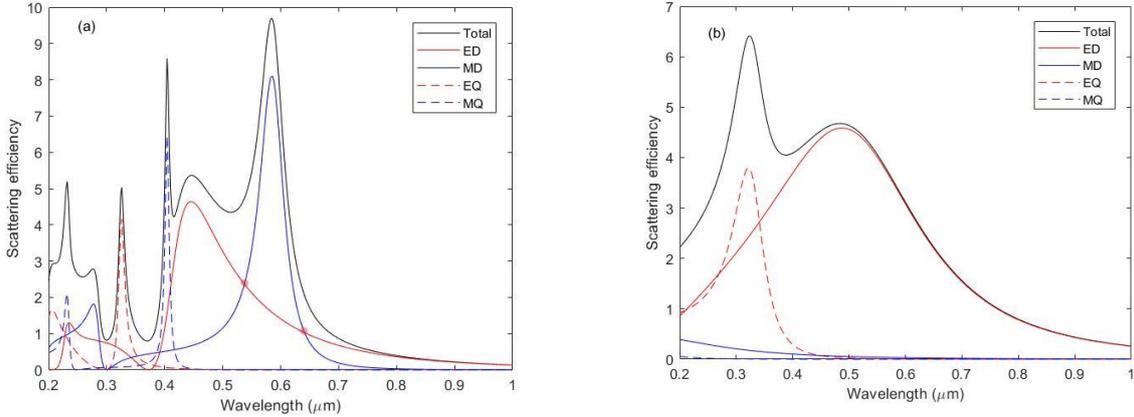

Figure 1. Scattering of the dielectric nanoparticle and Ag[16] nanosphere with radius r=80nm in the wavelength range of 0.2-1μm. (a) Scattering of the dielectric nanoparticle, including ED, MD, EQ, MQ, respectively. (b) Scattering of Ag nanoparticle, including ED, MD, EQ, MQ, respectively.

From figure 1(a), the dielectric nanoparticle has ED, MD, EQ, MQ, all resonance peaks, and its MD resonance is corresponding to its high real refractive index and satisfies the condition $\frac{\lambda_0}{n_0} = 2R_0$, where $\lambda_0$ is the free space wavelength, $R_0$ is the radius of the nanoparticle and $n_0$ is the refractive index. Also, an overlap of ED and MD mode is observed. The origin of electrical resonance for the dielectric nanoparticle is believed as the oscillation of bounded electrons, which originate displacement currents, and when the incident light frequency matches the natural frequency of bounded electrons' oscillation, its intensity is the strongest[17], and it can be resolved into the superposition of electrical multipoles; while the magnetic resonance occurs at a longer wavelength and a stronger intensity with the same origin. What's more, in two points in figure 1(a) (corresponding to wavelength 538 and 642nm, brightened), scattering efficiency of ED and MD are equal, which are considered as the points for Kerker condition. We will discuss it after Figure 2.

As figure 1(b) shows, for Ag nanoparticle, because Ag can emit free electrons, its electrical resonance is the strongest when incident light frequency meets its free electron oscillation frequency, while its magnetic resonance is highly suppressed. Only electric resonance works for Ag's scattering, and it has a wide peak for ED resonance and a narrow peak for EQ resonance. The high electric resonance efficiency is always used as surface plasmon for some strength-enhancement nanotechnologies such as nanoantenna[18] and surface-enhanced Raman scattering[19].

Figure 2 shows the far-field scattering graph with (a) λ=0.642μm and (b) λ=0.538μm, both are labeled in figure 1(a), (c) for the dielectric nanoparticle's MD and (d) for the Ag nanoparticle's ED.

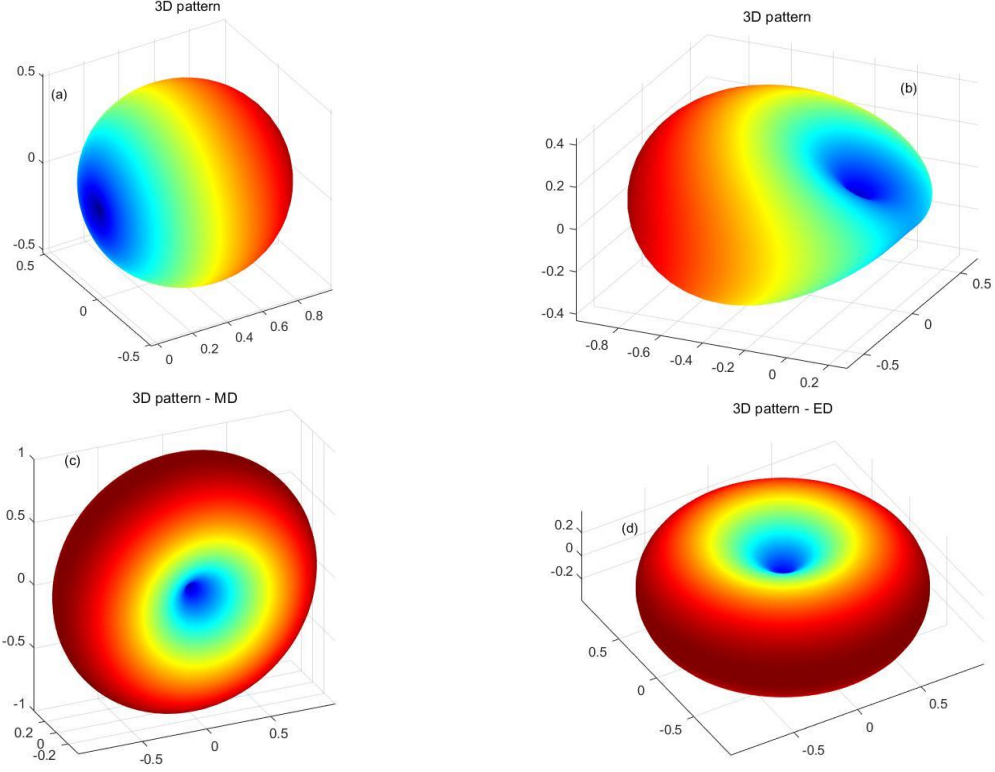

Figure 2. 3D pattern for the first and second Kerker conditions of the dielectric nanoparticle, the MD of the dielectric nanoparticle and the ED of the Ag nanoparticle. (a) λ=0.642μm. (b) λ=0.538μm. (c) MD of the dielectric nanoparticle. (d) ED of the Ag nanoparticle.

Fig.2(a) represents the first Kerker condition, which happens with indicating a strong forward scattering and a complete suppression of backward scattering, and this can be described as complete azimuthal symmetry[20]. Here, we only consider first-order resonance (ED and MD), so the total electric field is[21]

$$\boldsymbol{E}_{far}(\boldsymbol{r}) = \frac{k^2}{4\pi\varepsilon_0}p_x\frac{e^{ikr}}{r}(-sin\varphi\hat{\varphi} + cos\theta cos\ \varphi\hat{\theta}) \\ -\frac{Zk^2}{4\pi}m_y\frac{e^{ikr}}{r}\left(cos\ \theta\ sin\ \varphi\hat{\varphi} - cos\ \varphi\hat{\theta}\right) \quad (1)$$

where r, $\varphi$, $\theta$ are spherical coordinates, $p_x$ is electric dipole moment and $m_y$ is magnetic moment. Using Eq.1, the backward scattering efficiency is

$$\sigma_{back} = \lim_{r\to\infty} 4\pi r^2 \frac{\left|\boldsymbol{E}_{far}(\varphi=0,\theta=\pi)\right|^2}{|\boldsymbol{E}_{inc}|^2} \\ = \frac{k^4}{4\pi\epsilon^2|\boldsymbol{E}_{inc}|^2}\left|p_x - \frac{\sqrt{\epsilon_r}m_y}{c}\right|^2 \quad (2)$$

where $\boldsymbol{E}_{inc}$ is the incident electric field. From Eq.2, one can find that if $p_x -$

$\frac{\sqrt{\epsilon_r}m_y}{c} = 0$, then $\sigma_{back} = 0$, and this is the first Kerker condition[9].

While, fig.2(b) represents the second Kerker condition, showing a strong backward scattering and an incomplete suppression of forward scattering, derived from incomplete azimuthal symmetry[20], and leaves forward scattering efficiency to[12]

$$\frac{d\sigma_{sca}}{d\Omega}(0°) = \frac{16k^{10}}{9}\left|\frac{\alpha_e}{\epsilon_s}\right|^4 \qquad (3)$$

where $\alpha_e = p_x/(\epsilon_0|\boldsymbol{E}_{inc}|)$, $\epsilon_0$ and $\epsilon_s$ are the permittivity of the nanosphere and surrounding medium (vacuum for our research).

It is worth noting that such phenomena of vanishing backward or forwards scattering could also be realized at shorter wavelengths when quadrupole terms are considered, naming the generalized Kerker condition.

## Ag-Si core-shell nanoparticles

### Mode evolution of electric and magnetic dipolar resonances

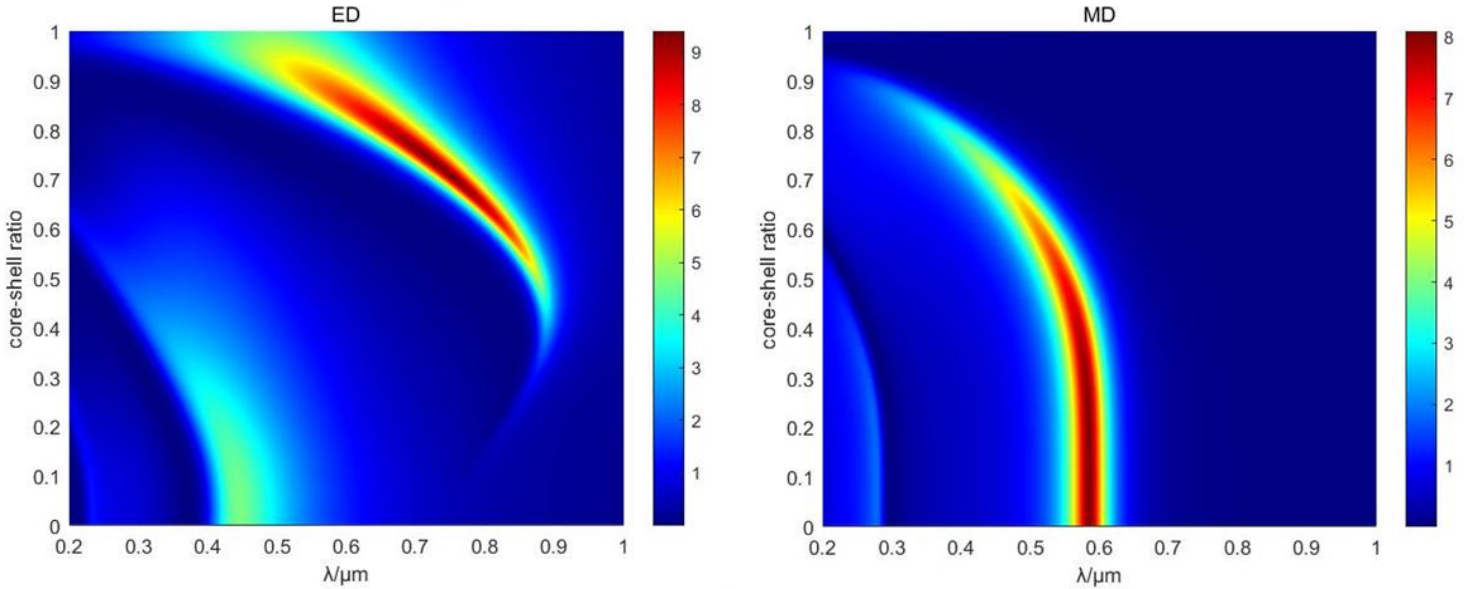

Figure 3. ED and MD resonances' scattering efficiency varied with wavelength and core-shell ratio.

To systematically view the relationship between ED, MD resonance, their corresponding wavelength and core-shell ratio, and continuously view the evolution of these resonances, we expand the method for single core-shell ratio's result to continuously core-shell ratio varied from 0 to 1. Here, core-shell ratio=0 means pure Si nanoparticle and core-shell ratio=1 means pure Ag nanoparticle, and we set the radius of this nanoparticle to 80nm, the same as homogenous sphere above. The refractive index for Si and Ag is also the same as above paragraph. Figure 3 shows ED and MD resonance coefficients varied with core-shell ratio and the wavelength of incident light.

For ED resonance, on the left side (wavelength 0.4-0.5μm), there is a zone indicating strong ED resonance. After comparing with figure 1, we can deduce that it is derived from Si. With the increasing of core-shell ratio, its wavelength and intensity are all decreasing, and finally vanish when core-shell ratio is over 0.5. On the right side, the wavelength of ED resonance first increases and then decreases with the increasing of core-shell ratio, and ED's scattering efficiency is nearly negligible when core-shell ratio is less than 0.4, but soon increases greatly and gets its peak when core-shell ratio is approximately 0.7, much stronger than ED resonance made by Si, and finally decreases with the core-shell ratio increases to 1. Moreover, the width of ED's peak also increases synchronously with core-shell ratio. We think this evolution progress is determined by the coupling of Ag and Si's ED resonance and the evolution of equivalent refractive index (imaginary part increase but real part increase at first and then decrease), and there has a best solution of ED's peak and the peak's width for different situations.

In detail, we find that ED resonance is mainly determined by Si when core-shell ratio is less than 0.4, but Si-determined mode quickly transits to Ag-determined mode----a high peak of ED----with the core-shell ratio increases from 0.4 to 0.5. To capture this process more precisely, Figure 4 shows the total scattering efficiency varied with core-shell ratio and wavelength, and Figure 5 shows the scattering efficiency with core-shell ratio equal to 0.4, 0.45, 0.5. At first, we note that the peak line in wavelength 0.4μm and the sharp peak in core-shell ratio 0.9-1 are not considered, because they correspond to MQ and EQ resonance.

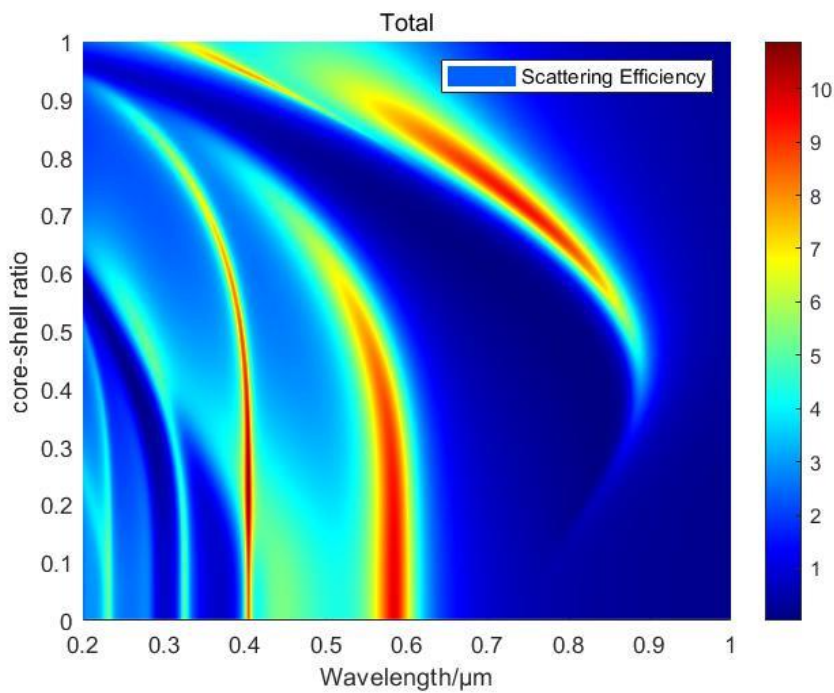

Figure 4. Total scattering efficiency varied with core-shell ratio and wavelength.

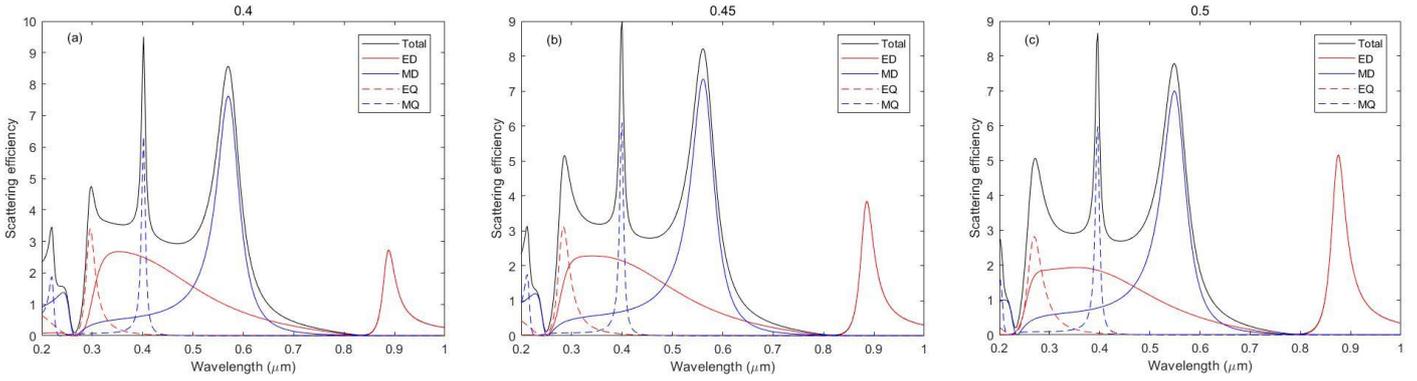

Figure 5. Scattering efficiency when core-shell ratio is (a) 0.4, (b) 0.45 and (c) 0.5.

From Figure 4-5, we find that when core-shell ratio increases from 0.4 to 0.5, ED that is made from Ag quickly increase while Si-determined MD remains stable, which means that in the core-shell ratio interval [0.4, 0.5], Si-determined mode (MD) does not replace by Ag-determined (ED) mode. They can co-exist. We think this co-existence can be used for differential scattering, which means we can control it's scattering mode (ED or MD) with wavelength and their efficiencies are approximately equal.

For MD resonance, MD resonance's wavelength and scattering efficiency peak remains stable when core-shell ratio is less than 0.4, but after that the wavelength and efficiency peak decrease and finally vanishes when core-shell ratio reaches over 0.9 (Figure 6). From the discussion of ED, we think for MD when core shell ratio is less than 0.4, Ag does not hold the position and $\frac{\lambda_0}{n_0} = 2R_0$ is approximately available, and then Ag's mode quickly evolutes into main point, due to the decreasing of Si shell, the imaginary part of the refractive index for this core-shell nanoparticle increase rapidly and suppresses MD.

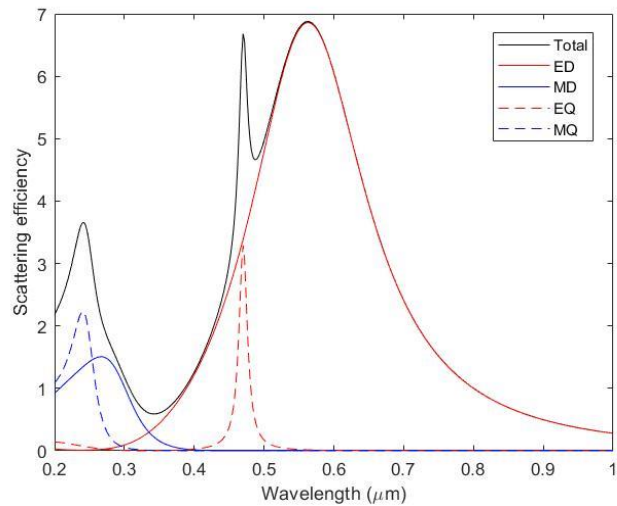

Figure 6. Scattering efficiency when core-shell ratio is 0.9.

## Tuning scattering directionality via varying core-shell ratio

Figure 7 shows Kerker conditions varied with core-shell ratio and incident light's wavelength. We use ED and MD's scattering efficiency as our first criterion: if they are equal and core-shell ratio is not too big, then the points meet our criterion satisfied for Kerker conditions. (For each result as core shell ratio increased from 0, 0.1 to 0.8, it really corresponds to its Kerker condition.) To check, we would find imag($a_1$) and imag($b_1$) ($a_1$ and $b_1$ are the first-order electrical and magnetic scattering efficiency) corresponding to these points. If they are equal or opposite, then these points satisfy Kerker conditions.

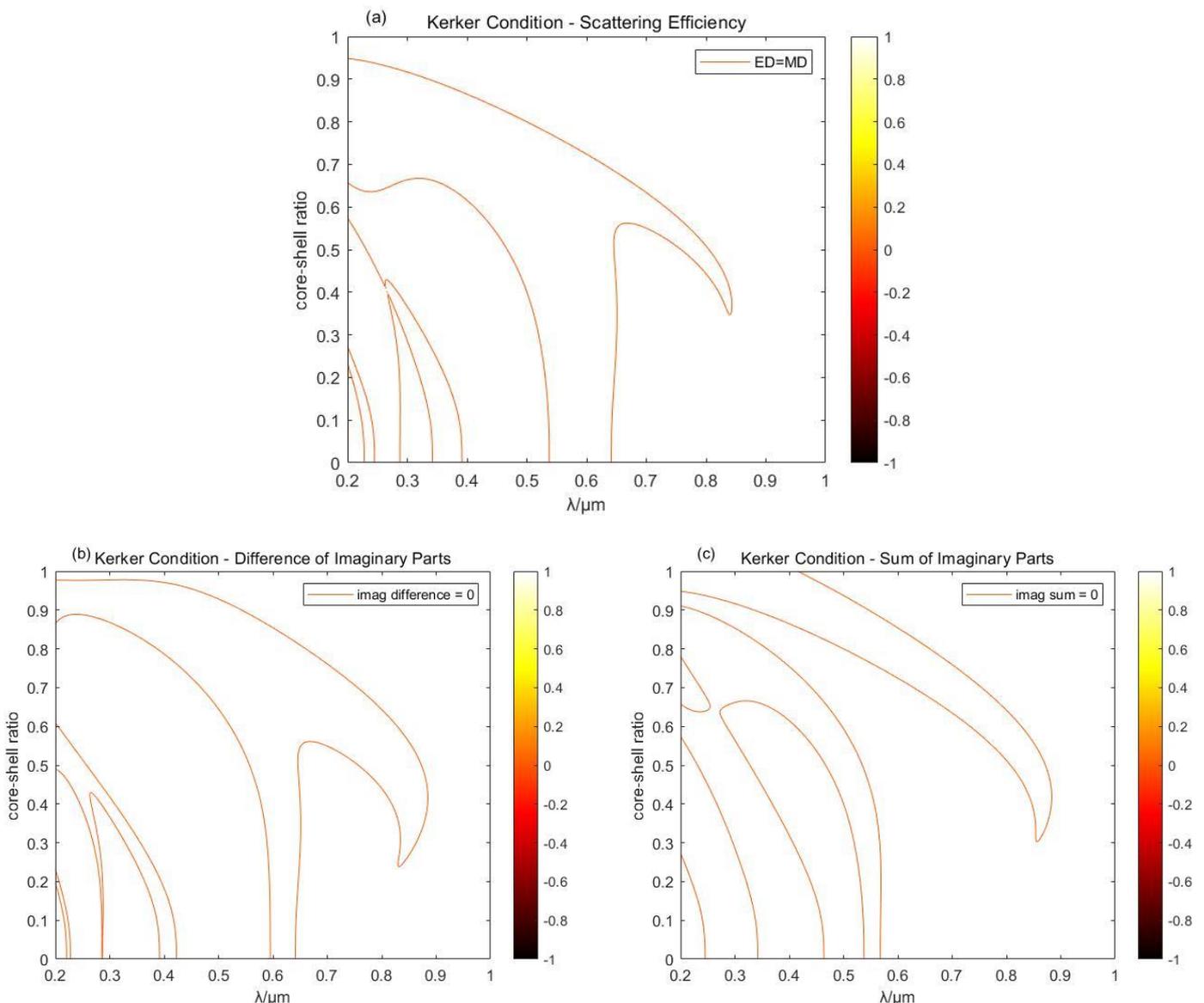

Figure 7. Kerker conditions varied with core-shell ratio and incident light's wavelength. (a) the difference of ED and MD's scattering efficiency; (b) the difference of the imaginary part of $a_1$ and $b_1$; (c) the sum of the imaginary part of $a_1$ and $b_1$.

From figure 7(a), we can determine the wavelength that matches Kerker condition is close to 540 and 640nm with core-shell ratio less than 0.7, corresponding to the result of Figure 1(b) and 2, and its evolution is very similar to MD. Both figure 7(b) and 7(c) shows a similar line starting from λ=600nm and evolutes like MD resonance, but after checking with figure 7(a), we can find that this wavelength are not suit for Kerker conditions----the suitable wavelength is about 640 and 540nm, corresponding to figure 7(b) and 7(c), respectively. From figure 2, we suppose that 7(b) shows the first Kerker condition and 7(c) shows the second Kerker condition. To check our conjecture, we first intermittently find Kerker conditions for core-shell ratio increases from 0.1 to 0.9, step by step with stride 0.1, and we draw all these results' far-field graph (see in figure 8) with table 1 (see in Appendix) showing the results of Kerker conditions in our criteria.

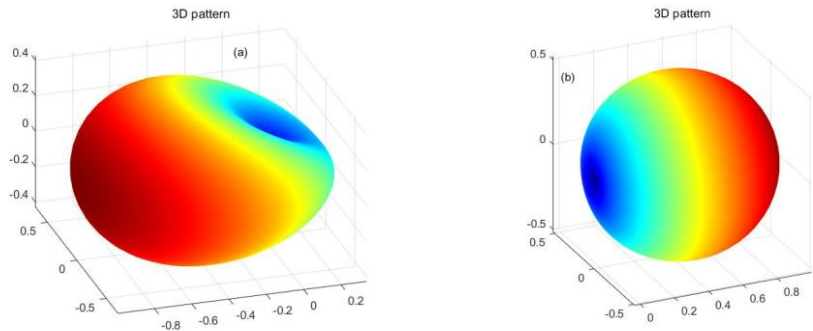

Figure 8 (a) and (b), core-shell ratio=0.1, wavelength=536 and 642nm;

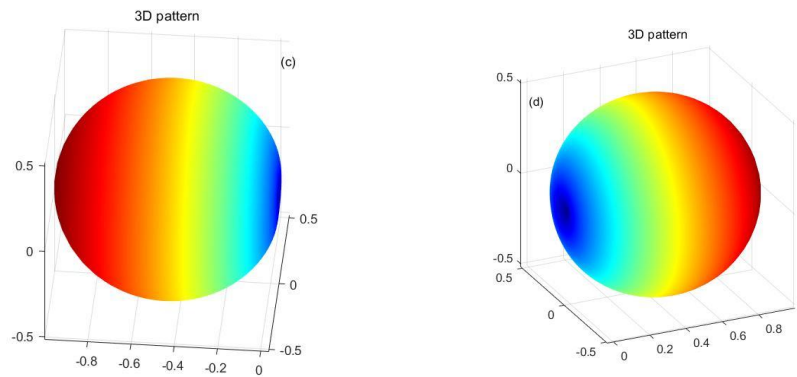

Figure 8 (c) and (d), core-shell ratio=0.2, wavelength=532 and 646nm;

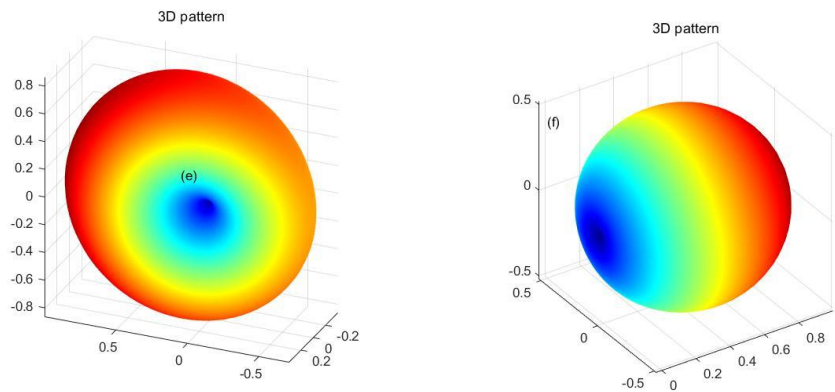

Figure 8 (e) and (f), core-shell ratio=0.3, wavelength=520 and 650nm;

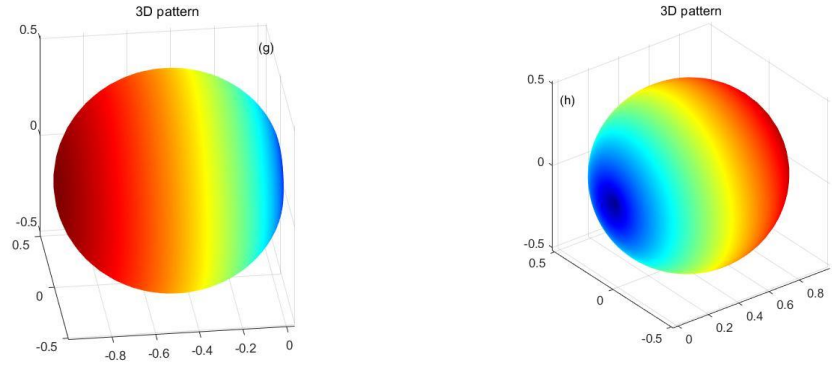

Figure 8 (g) and (h), core-shell ratio=0.4, wavelength=500 and 650nm;

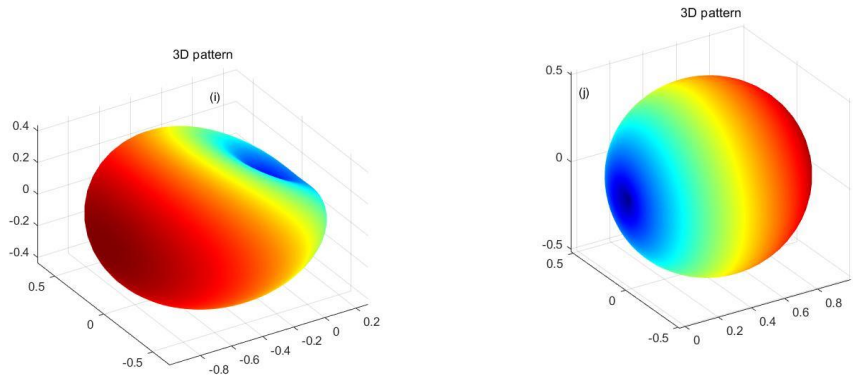

Figure 8 (i) and (j), core-shell ratio=0.5, wavelength=468 and 644nm;

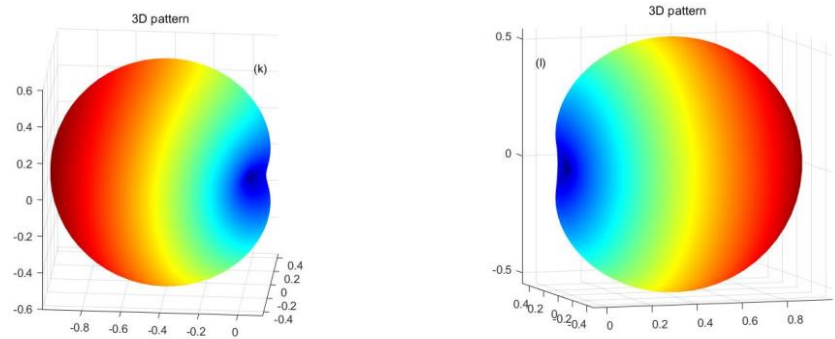

Figure 8 (k) and (l), Core-shell ratio=0.6, wavelength=412 and 648nm;

From figure 8, we find that with the increasing of core-shell ratio, the first Kerker condition's wavelength remains stable with small changes, and approximately matches $\frac{\lambda_0}{n_0} = 2R_0$, while the second Kerker condition's wavelength first keep stable and then quickly decrease, its evolution really like MD resonance (figure 9 (a)). They correspond to figure 7 (a) exactly. However, for scattering intensity, the first Kerker condition's intensity gets weaker continuously, while the second Kerker condition's intensity first gets weaker and then stronger (figure 9 (b)). For core-shell ratio more than 0.7, we can't find Kerker condition because Si's ED resonance is too weak to compare with total's MD or Ag's ED resonance, the same as figure 7 (a) shows.

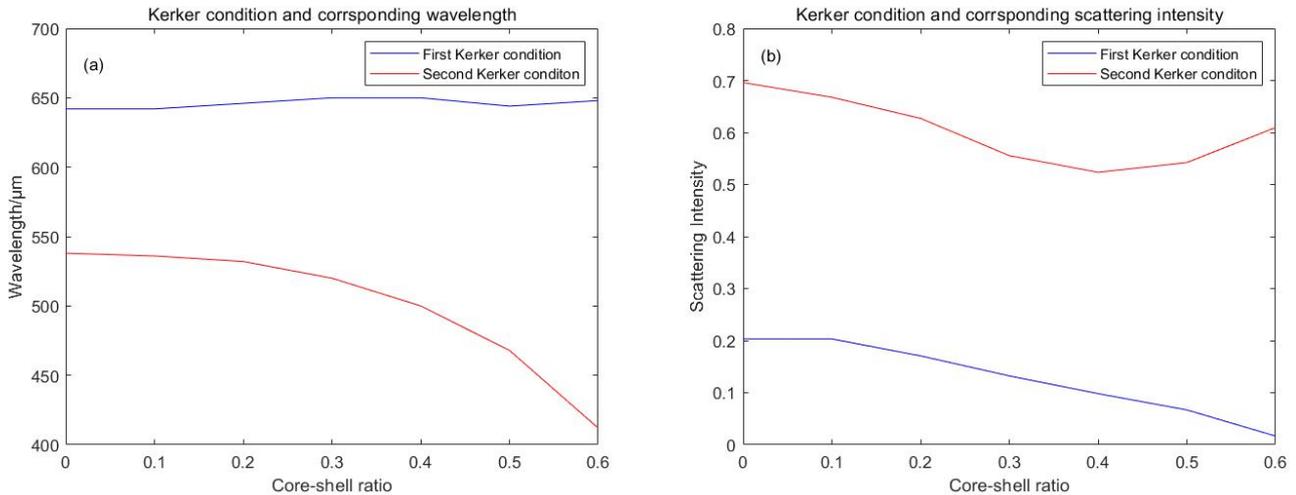

Figure 9. Wavelength and scattering intensity for the first and second Kerker conditions varied with core-shell ratio. (a) wavelength. (b) scattering intensity.

While, there are also something that does not match figure 7 in figure 8. For example, figure 3 (e) is not a far-field graph of Kerker condition, but the wavelength and scattering coefficients match our criterion, indicating that high-order scattering participates and couples with it; we first suppose all short-wavelength Kerker conditions are the second kind and all the others are first kind, and their far-field graph should be similar, respectively, but they only show opposite direction, and graph's shape does not always match figure 2 and Eq.2-3. For example, as figure 8 (g) and (h) show, sometimes a nearly completely suppression of inverse scattering happens in the both sides, not always like figure 2----a complete suppression of backward scattering (figure 2(b)) and an incomplete suppression of forward scattering (figure 2(a)).

# Conclusion

In this research, we demonstrate the evolution of ED, MD resonance and unidirectional scattering with different core-shell ratio in a single Ag-Si nanoparticle, and we find that ED resonance is mainly determined by Ag and MD resonance is completely determined by Si. They couple with each other, switch the mode in a short range and both their properties can appear in a single nanoparticle, like high ED resonance with unidirectional scattering in different wavelengths and directions. This means we can control the scattering properties of a core-shell nanoparticle by adjusting the core-shell ratio, which has a further interest in nanotechnology with high-level machining.

## Methods

We use these formulas derived from Mie theory to calculate Mie scattering coefficients $a_n$ and $b_n$ for pure nanosphere[7]:

$$a_n = \frac{m\psi_n(mx)\psi_n'(x) - \psi_n(x)\psi_n'(mx)}{m\psi_n(mx)\xi_n'(x) - \xi_n(x)\psi_n'(mx)}$$
$$b_n = \frac{\psi_n(mx)\psi_n'(x) - m\psi_n(x)\psi_n'(mx)}{\psi_n(mx)\xi_n'(x) - m\xi_n(x)\psi_n'(mx)}$$
(4)

And these to calculate Mie scattering coefficients $a_n$ and $b_n$ for core-shell nanosphere:

$$a_n = \frac{\psi_n(y)[\psi_n'(m_2 y) - A_n \chi_n'(m_2 y)] - m_2 \psi_n'(y)[\psi_n(m_2 y) - A_n \chi_n(m_2 y)]}{\xi_n(y)[\psi_n'(m_2 y) - A_n \chi_n'(m_2 y)] - m_2 \xi_n'(y)[\psi_n(m_2 y) - A_n \chi_n(m_2 y)]}$$
$$b_n = \frac{m_2 \psi_n(y)[\psi_n'(m_2 y) - B_n \chi_n'(m_2 y)] - \psi_n'(y)[\psi_n(m_2 y) - B_n \chi_n(m_2 y)]}{m_2 \xi_n(y)[\psi_n'(m_2 y) - B_n \chi_n'(m_2 y)] - \xi_n'(y)[\psi_n(m_2 y) - B_n \chi_n(m_2 y)]}$$
$$A_n = \frac{m_2 \psi_n(m_2 x)\psi_n'(m_1 x) - m_1 \psi_n'(m_2 x)\psi_n(m_1 x)}{m_2 \chi_n(m_2 x)\psi_n'(m_1 x) - m_1 \chi_n'(m_2 x)\psi_n(m_1 x)}$$
$$B_n = \frac{m_2 \psi_n(m_1 x)\psi_n'(m_2 x) - m_1 \psi_n(m_2 x)\psi_n'(m_1 x)}{m_2 \chi_n'(m_2 x)\psi_n(m_1 x) - m_1 \psi_n'(m_1 x)\chi_n(m_2 x)}$$
(5)

where $x=kR_1$, $y=kR_2$, $k=2\pi/\lambda$ ($\lambda$ is the incident light wavelength in free space), $R_1$ is core radius and $R_2$ is the radius of nanosphere; m is the refractive index for a pure sphere, $m_1$ is the refractive index of core and $m_2$ is the refractive index of shell; $\psi_n$ is spherical Bessel function of the first kind and $\psi_n'$ is its derivative function, $\xi_n$ is the spherical Hankel function of the first kind and $\xi_n'$ is its derivative function, $\chi_n$ is the spherical Bessel function of the second kind and $\chi_n'$ is its derivative function, n is positive internal, represents $2^n$ electric or magnetic multipoles.

# Appendix

Table 1. Kerker conditions varied with core-shell ratio (for figure 8).

| Core-shell ratio | Wavelength/nm | Re($a_1$) | Im($a_1$) | Im($b_1$) |
|---|---|---|---|---|
| 0 | 538 | 0.348 | -0.4763 | 0.4763 |
| | 642 | 0.1017 | -0.3022 | |
| 0.1 | 536 | 0.3338 | -0.4716 | 0.4716 |
| | 642 | 0.1016 | -0.3021 | |
| 0.2 | 532 | 0.3136 | -0.4639 | 0.4639 |
| | 646 | 0.08539 | -0.2794 | |
| 0.3 | 520 | 0.278 | -0.4477 | 0.4477 |

| | 650 | 0.06635 | -0.2484 | |
|---|---|---|---|---|
| 0.4 | 500 | 0.2624 | -0.4392 | 0.4392 |
| | 650 | 0.04963 | -0.2159 | |
| 0.5 | 468 | 0.2721 | -0.4439 | 0.4439 |
| | 644 | 0.03433 | -0.1796 | |
| 0.6 | 412 | 0.3059 | -0.4596 | 0.4596 |
| | 648 | 0.01348 | -0.09014 | -0.1114 |
| >0.7 | No wavelength suits Kerker conditions. | | | |